\documentclass[preprint, 10pt, 5p]{elsarticle}
\usepackage[utf8]{inputenc}
\usepackage{amsmath}
\usepackage{empheq}
\usepackage{amsfonts}
\usepackage{amssymb}
\usepackage{mathrsfs}
\usepackage{subfigure}
\usepackage[hidelinks]{hyperref}
\usepackage{outlines}
\usepackage{tikz}
\usepackage{booktabs}

\newcommand{\br}{\mathbf{r}}

\newcommand{\bv}{\mathbf{v}}

\newcommand{\bB}{\mathbf{B}}
\newcommand{\bF}{\mathbf{F}}

\newcommand{\D}{\mathrm{d}}

\newcommand{\bk}{\boldsymbol{\kappa}}
\newcommand{\bz}{\boldsymbol{\zeta}}

\begin{document}
	
\biboptions{sort&compress}
	
\title{Extending the Single-Fluid Solvability Conditions for More General Plasma Systems}
\date{\today}
\author{E.~J.~Kolmes}
\ead{ekolmes@princeton.edu}
\address{Department of Astrophysical Sciences, Princeton University, Princeton, New Jersey 08544, USA}
\author{M.~E.~Mlodik}
\address{Department of Astrophysical Sciences, Princeton University, Princeton, New Jersey 08544, USA}
\author{N.~J.~Fisch}
\address{Department of Astrophysical Sciences, Princeton University, Princeton, New Jersey 08544, USA}

\begin{abstract}
We extend the single-fluid solvability conditions to include plasma systems at arbitrary $\beta$ with arbitrary flows and external forcing terms. This treatment includes both the isotropic and the anisotropic cases. The generalized conditions that result can be used to generate certain classes of single-fluid equilibria. 
\end{abstract}
	
\maketitle
	
\section{Introduction}

In 1963, Taylor studied \cite{Taylor1963} the ideal MHD and anisotropic MHD solvability conditions for arbitrary magnetic configurations. 
In the ideal MHD case, Taylor found two conditions which the magnetic field and pressure profiles must satisfy in equilibrium: 
\begin{gather}
\hat b \cdot \nabla p = 0 \label{eqn:idealMHDParallel} \\
\oint \frac{\nabla B \cdot (\nabla p \times \bB)}{B^4} \, \D s = 0 . \label{eqn:idealMHDPerp}
\end{gather}
Here $p$ is the (scalar) pressure, $\bB$ is the magnetic field, $B$ is the magnitude of $\bB$, $\hat b \doteq \bB / B$, and the integral in Eq.~(\ref{eqn:idealMHDPerp}) is taken along some closed field line, with $s$ being a field-aligned coordinate. 
This result does not require any assumptions on the size of $\beta$ (that is, the ratio of the plasma pressure to the magnetic pressure). 
If a field line only contains plasma between some $s_0$ and $s_1$, then Eq.~(\ref{eqn:idealMHDPerp}) can be rewritten as 
\begin{gather}
\int_{s_0}^{s_1} \frac{\nabla B \cdot (\nabla p \times \bB)}{B^4} \, \D s = 0 
\end{gather}
so long as the parallel component of the current $\mathbf{j}$ vanishes at either end of the plasma. 
The second condition is related to work done on the general theory of what are sometimes called ``magnetic differential equations" \cite{Kruskal1958, Newcomb1959}.

Taylor also calculated a corresponding set of conditions for the case in which the pressure is anisotropic and $\beta$ is small. 
If the pressure tensor $P$ is given by 
\begin{gather}
P = p_\perp \mathbf{I} + (p_{||} - p_\perp) \hat b \hat b, \label{eqn:CGLPressure}
\end{gather}
where $\mathbf{I}$ is the unit tensor and $p_{||}$ and $p_\perp$ are the parallel and perpendicular pressures, respectively, then the two conditions become 
\begin{gather}
\hat b \cdot \nabla p_{||} + \frac{p_\perp-p_{||}}{B} \, \hat b \cdot \nabla B = 0 \label{eqn:CGLParallel} \\
\oint \frac{\nabla (p_{||}+p_\perp)}{B^4} \cdot (\bB \times \nabla B) \, \D s = 0 . \label{eqn:CGLPerpLowBeta}
\end{gather}
The small-$\beta$ assumption is used only in the derivation of Eq.~(\ref{eqn:CGLPerpLowBeta}). 

Some years later, Hall and McNamara used a guiding-center fluid model to derive a finite-$\beta$ analog to Eq.~(\ref{eqn:CGLPerpLowBeta}) \cite{Hall1975}. 
In particular, if $\bk \doteq \hat b \cdot \nabla \hat b$, they found that Eq.~(\ref{eqn:CGLPerpLowBeta}) becomes 
\begin{gather}
\oint \frac{\nabla (p_{||} + p_\perp)}{B^3} \cdot (\bB \times \bk) \, \D s = 0. \label{eqn:CGLPerpFiniteBeta}
\end{gather}
Eq.~(\ref{eqn:CGLPerpFiniteBeta}) has the same symmetry properties as Eqs.~(\ref{eqn:idealMHDPerp}) and (\ref{eqn:CGLPerpLowBeta}). 

In the low-$\beta$ limit, $\bB$ can be fixed independently of $P$, and Taylor showed that these solvability conditions -- that is, Eqs.~(\ref{eqn:idealMHDParallel}) and (\ref{eqn:idealMHDPerp}) for the isotropic case and Eqs.~(\ref{eqn:CGLParallel}) and (\ref{eqn:CGLPerpLowBeta}) for the anisotropic case -- are necessary and sufficient conditions for a given $P$ to be a solution for a given nonvanishing $\bB$. 
At higher $\beta$, the situation is more complicated: the solvability conditions must still be satisfied, but $\bB$ must also itself be consistent with the plasma currents. 

These solvability conditions are interesting for three main reasons. 
First: they provide general statements about the nature of all possible equilibria. 
Second: they can be used to generate special families of equilibria. 
Third: these special families have often turned out to have very good stability properties, and are closely related to the origins of the theory of omnigeneity \cite{Hall1975, Post1987}. 

The original work on these solvability conditions considered only static equilibria. 
However, there are many systems of interest for equilibria that include flows. 
For example, centrifugal confinement devices (rotating mirrors) rely on the centrifugal forces from rotation to achieve longitudinal confinement \cite{Lehnert1971, Bekhtenev1980, Ellis2001, Ellis2005, Fetterman2008, Teodorescu2010, Agren2017, White2018, Kolmes2021HotIon, Rubin2021}. 
It is possible to design a toroidal confinement device in which poloidal flows provide the rotational transform, replacing the role of the poloidal field in a tokamak \cite{Rax2017, Ochs2017ii}.
There are also a number of fusion devices, including stellarators and tokamaks, for which flows are not an element of the basic design but which develop (or can be induced to develop) equilibrium flow structures \cite{Suckewer1979, Helander2008, deGrassie2009, Estrada2009, Yan2010, StoltzfusDueck2012, Cvejic2022}. 
Moreover, rotation is centrally important for a number of mass filter concepts \cite{Lehnert1971, Krishnan1981, Krishnan1983, Fetterman2011b, Gueroult2012MCMF, Rax2016, Dolgolenko2017, Zweben2018}. 
Despite all of this, the analytic theory for single-fluid equilibria with steady-state flows is comparatively underdeveloped. 

There has been some discussion in the literature regarding the extension of the solvability conditions to more general systems. 
Tessarotto \textit{et al.} discuss these solvability conditions for a flowing system, though their focus is not on the explicit evaluation of these expressions, and they leave the analog of Eq.~(\ref{eqn:CGLPerpFiniteBeta}) in terms of undetermined coefficients \cite{Tessarotto1997}. 
Kotelnikov and Rom\'e derive an extension of Taylor's isotropic result for a nonneutral plasma \cite{Kotelnikov2008}; they considered the slow-rotation limit, so their extension includes an electrostatic potential but not the inertial effects of the flow itself. 
Their result has applications in the study of Penning-Malmberg trap equilibria \cite{Kotelnikov2008, Akhmetov2009, Rome2010}. 

This paper is organized as follows. 
Section~\ref{sec:generalization} generalizes the solvability conditions to allow for cases with steady-state flows and arbitrary external forces. 
Along the way, it rederives Eq.~(\ref{eqn:CGLPerpFiniteBeta}) without the need for any of the additional assumptions made by Hall and McNamara. 
Section~\ref{sec:solutions} describes families of equilibria for certain special cases. 
Section~\ref{sec:isodynamicity} explores the relationship of these results with isodynamicity. 
Section~\ref{sec:conclusion} is a discussion of these results. 

\section{Generalization of Solvability Conditions} \label{sec:generalization}

Consider a fluid model with 
\begin{gather}
0 = \mathbf{j} \times \bB - \nabla \cdot P + \bz , \label{eqn:CGLForceBalance}
\end{gather}
where $\bz$ is an arbitrary vector and $P$ is given by Eq.~(\ref{eqn:CGLPressure}). 
The $\bz$ term could represent an external force term (gravity, for example); if $\rho$ is the mass density and $\bF$ is the force, this would be $\bz = \rho \bF$. This term can also capture the effects of a steady-state velocity $\bv$ (for example, centrifugal forces) by setting $\bz = - \rho \bv \cdot \nabla \bv$. 

The divergence of $P$ is 
\begin{align}
\nabla \cdot P &= \nabla p_\perp + \big[ \hat b \cdot \nabla (p_{||}-p_\perp) + (p_{||}-p_\perp) \nabla \cdot \hat b \big] \hat b \nonumber \\
&\hspace{75 pt}+ (p_{||}-p_\perp) \bk, 
\end{align}
where, as before, $\bk \doteq \hat b \cdot \nabla \hat b$. 
Then the $\hat b$ component of Eq.~(\ref{eqn:CGLForceBalance}) is 
\begin{gather}
\hat b \cdot \nabla p_{||} - \frac{p_{||}-p_\perp}{B} \, \hat b \cdot \nabla B = \hat b \cdot \bz . \label{eqn:generalParallelCondition}
\end{gather}
Eq.~(\ref{eqn:generalParallelCondition}) is the generalization of Eq.~(\ref{eqn:CGLParallel}). 

The second solvability condition is somewhat more involved. 
Following Taylor, note that in steady state Maxwell's equations require 
\begin{gather} 
\nabla \cdot \mathbf{j} = 0. 
\end{gather}
Let $\mathbf{j}_{||} \doteq \hat b \hat b \cdot \mathbf{j}$ and $\mathbf{j}_\perp \doteq (\mathbf{I} - \hat b \hat b) \mathbf{j}$. 
There must exist some scalar function $\lambda(\br)$ such that $\mathbf{j}_{||} = \lambda \bB$. 
Then 
\begin{align}
\nabla \cdot \mathbf{j} = \bB \cdot \nabla \lambda + \nabla \cdot \mathbf{j}_\perp , 
\end{align}
which implies that 
\begin{gather}
\oint \frac{\nabla \cdot \mathbf{j}_\perp}{B} \, \D s = 0, 
\end{gather}
where, as before, the integral is taken over a field line. 
This is how Taylor derived Eqs.~(\ref{eqn:idealMHDPerp}) and (\ref{eqn:CGLPerpLowBeta}). 
The difficulty lies in computing $\nabla \cdot \mathbf{j}_\perp$. 
Taylor simplified the calculation in the anisotropic case by letting $\beta \rightarrow 0$, but it is possible to compute $\nabla \cdot \mathbf{j}_\perp$ without doing so. 

With that in mind, note that Eq.~(\ref{eqn:CGLForceBalance}) implies that 
\begin{gather}
\mathbf{j}_\perp = - \frac{(\nabla \cdot P - \bz) \times \bB}{B^2} \, .
\end{gather}
The divergence of $\mathbf{j}_\perp$ can be computed as follows: 
\begin{align}
\nabla \cdot \mathbf{j}_\perp &= - \bB \cdot \nabla \times \bigg( \frac{\nabla \cdot P - \bz}{B^2} \bigg) + \frac{\nabla \cdot P - \bz}{B^2} \cdot \mu_0 \mathbf{j} . 
\end{align}
Eq.~(\ref{eqn:CGLForceBalance}) implies that the last term must vanish, so 
\begin{align}
\nabla \cdot \mathbf{j}_\perp 
&= \frac{\bB \cdot \nabla \times \bz}{B^2} + \frac{2 (\nabla \cdot P - \bz)}{B^3} \cdot (\bB \times \nabla B) \nonumber \\
&\hspace{50 pt}- \frac{\bB}{B^2} \cdot \nabla \times (\nabla \cdot P) . \label{eqn:intermediate1}
\end{align}
The second and third terms can be simplified individually. First, 
\begin{align}
&\frac{2 (\nabla \cdot P - \bz)}{B^3} \cdot (\bB \times \nabla B) \nonumber \\
&\hspace{10 pt}= \frac{2 \big[ \nabla p_\perp + (p_{||}-p_\perp) \bk - \bz \big]}{B^2} \cdot \bigg( \bB \times \frac{\nabla B}{B} \bigg) . 
\end{align}
Note that 
\begin{align}
\bk &= \frac{\mu_0 \mathbf{j} \times \bB}{B^2} + \bigg(\mathbf{I}-\hat b \hat b \bigg) \cdot \frac{\nabla B}{B} \, , \label{eqn:kappa}
\end{align}
and note that 
\begin{align}
\big[ \nabla p_\perp + (p_{||}-p_\perp) \bk - \bz \big] \cdot \mathbf{j}_\perp &= ( \nabla \cdot P - \bz ) \cdot \mathbf{j}_\perp \nonumber \\
&=0 , 
\end{align}
so 
\begin{align}
\frac{2 (\nabla \cdot P - \bz)}{B^3} \cdot (\bB \times \nabla B) 
&= \frac{2 (\nabla p_\perp-\bz)}{B^2} \cdot (\bB \times \bk). 
\end{align}
Meanwhile, the third term on the RHS of Eq.~(\ref{eqn:intermediate1}) can be simplified as follows: 
\begin{align}
&
- \frac{\bB}{B^2} \cdot \nabla \times (\nabla \cdot P)
 \nonumber \\
&\hspace{5 pt}= - \frac{\bB}{B^2} \cdot \bigg\{ \big[ \hat b \cdot \nabla (p_{||}-p_\perp) + (p_{||}-p_\perp) \nabla \cdot \hat b \big] \nabla \times \hat b \nonumber \\
&\hspace{80 pt}+ \nabla \times \big[ (p_{||}-p_\perp) \bk \big] \bigg\} \nonumber \\
&\hspace{5 pt}= - \bigg[ \bB \cdot \nabla \bigg( \frac{p_{||}-p_\perp}{B} \bigg) \bigg] \frac{\bB \cdot \mu_0 \mathbf{j}}{B^3} \nonumber \\
&\hspace{50 pt}- \frac{\bB}{B^2} \cdot \big[ \nabla (p_{||}-p_\perp) \times \bk \big] \nonumber \\
&\hspace{50 pt}- \frac{p_{||}-p_\perp}{B} \, \frac{\bB}{B} \cdot \nabla \times \bk . 
\end{align}
Then the divergence of $\mathbf{j}_\perp$ can be written as 
\begin{align}
&\nabla \cdot \mathbf{j}_\perp \nonumber \\
&= \frac{\bB \cdot \nabla \times \bz}{B^2} + \frac{\nabla (p_{||}+p_\perp)-2 \bz}{B^2} \cdot (\bB \times \bk) \nonumber \\
&\hspace{15 pt}- \bB \cdot \nabla \bigg( \frac{p_{||}-p_\perp}{B} \bigg) \frac{\mu_0 \lambda}{B} - \frac{p_{||}-p_\perp}{B} \frac{\bB}{B} \cdot \nabla \times \bk . 
\end{align}
It is now necessary to simplify the last term. 
To begin, 
\begin{align}
&\bB \cdot \nabla \times \bk \nonumber \\
&= \nabla \cdot ( \bk \times \bB ) + \bk \cdot \mu_0 \mathbf{j} \nonumber \\
&= \nabla \cdot \bigg[ \bigg( \frac{\mu_0 \mathbf{j} \times \bB}{B^2} + \frac{\nabla B}{B} \bigg) \times \bB \bigg] + \frac{\nabla B \cdot \mu_0 \mathbf{j}_\perp}{B} \nonumber \\
&= - \nabla \cdot (\mu_0 \mathbf{j}_\perp) - \frac{\nabla B \cdot \mu_0 \mathbf{j}}{B} + \frac{\nabla B \cdot \mu_0 \mathbf{j}_\perp}{B} \, .
\end{align}
Invoking the requirement that $\nabla \cdot \mathbf{j} = 0$, this is 
\begin{align}
\bB \cdot \nabla \times \bk &= \mu_0 \bB \cdot \nabla \lambda - \frac{\nabla B \cdot \mu_0 \lambda \bB}{B} \nonumber \\
&= \mu_0 B \bB \cdot \nabla \bigg( \frac{\lambda}{B} \bigg) \, . 
\end{align}
Then 
\begin{align}
\nabla \cdot \mathbf{j}_\perp &= \frac{\bB\cdot\nabla\times\bz}{B^2} + \frac{\nabla (p_{||}+p_\perp)-2 \bz}{B^2} \cdot (\bB \times \bk) \nonumber \\
&\hspace{70 pt}- \bB \cdot \nabla \bigg[ \frac{(p_{||}-p_\perp) \mu_0 \lambda}{B^2} \bigg] . \label{eqn:divJPerp}
\end{align}
As a result, the solvability condition is 
\begin{empheq}[box=\fbox]{align}
\oint \bigg[ \frac{\nabla (p_{||}+p_\perp)-2 \bz}{B^3} \cdot (\bB \times \bk) + \frac{\bB \cdot \nabla \times \bz}{B^3} \bigg] \, \D s = 0. \label{eqn:solvability} 
\end{empheq}
The above equation is the primary result of this section. 
The previous versions of the second solvability condition are all limits of Eq.~(\ref{eqn:solvability}). 

Hall and McNamara's finite-$\beta$ condition -- that is, Eq.~(\ref{eqn:CGLPerpFiniteBeta}) -- follows immediately from Eq.~(\ref{eqn:solvability}) by setting $\bz = 0$. 
Taylor's Eq.~(\ref{eqn:CGLPerpLowBeta}) then follows by noting that 
\begin{align}
\bB \times \bk &= \bB \times \bigg( \frac{\mu_0 \mathbf{j}\times\bB}{B^2} + \frac{\nabla B}{B} \bigg) \nonumber \\
&= \bB \times \bigg( \frac{\mu_0 (\nabla \cdot P - \bz)}{B^2} + \frac{\nabla B}{B} \bigg), 
\end{align}
so that the low-$\beta$ (small plasma pressure) limit is 
\begin{align}
\lim_{\beta \rightarrow 0} \bB \times \bk &= \bB \times \bigg( \frac{\nabla B}{B} - \frac{\mu_0 \bz}{B^2} \bigg). 
\end{align}
When $\bz = 0$, this low-$\beta$ limit yields Eq.~(\ref{eqn:CGLPerpLowBeta}). 
To retrieve the ideal MHD solvability condition, start by setting $p_{||}=p_\perp = p$: 
\begin{align}
\oint \bigg[ \frac{\nabla p -\bz}{B^3} \cdot (\bB \times \bk) + \frac{\bB \cdot \nabla \times \bz}{2 B^3} \bigg] \, \D s = 0. 
\end{align}
It follows from Eqs.~(\ref{eqn:CGLForceBalance}) and (\ref{eqn:kappa}) that this can be written equivalently as 
\begin{empheq}[box=\fbox]{align}
\oint \bigg[ \frac{\nabla p - \bz}{B^4} \cdot (\bB \times \nabla B) + \frac{\bB \cdot \nabla \times \bz}{2 B^3} \bigg] \, \D s = 0, 
\end{empheq}
which reduces to Eq.~(\ref{eqn:idealMHDPerp}) when $\bz = 0$, and is equivalent to the condition found by Kotelnikov and Rom\'e when $\bz = - n e \nabla \phi$ \cite{Kotelnikov2008}. 

\section{Special Solutions} \label{sec:solutions}

One special class of profiles that solve Eq.~(\ref{eqn:solvability}) are those for which the integrand itself vanishes at every point. 
Interestingly, the calculation in Section~\ref{sec:generalization} suggests that there are two possible formulations of this condition, depending on whether the last term in Eq.~(\ref{eqn:divJPerp}) is retained in the integral. 
If the last term is not kept, then the condition is 
\begin{gather}
\frac{\nabla (p_{||}+p_\perp) - 2 \bz}{B^3} \cdot (\bB \times \bk) + \frac{\bB \cdot \nabla \times \bz}{B^3} = 0. \label{eqn:integrandVanishes}
\end{gather}
If it is kept, the condition is instead 
\begin{align}
&\frac{\nabla (p_{||}+p_\perp) - 2 \bz}{B^3} \cdot (\bB \times \bk) + \frac{\bB \cdot \nabla \times \bz}{B^3} \nonumber \\
&\hspace{80 pt}- \frac{\bB}{B} \cdot \nabla \bigg[ \frac{(p_{||}-p_\perp) \mu_0 \lambda}{B^2} \bigg] = 0. \label{eqn:integrandVanishesWithDivergence}
\end{align}
As was true in the previous versions of this problem, there are two special cases of these profiles worth pointing out. 

\subsection{Symmetric Systems}

First, consider a system that is cylindrically symmetric, with radial, axial, and azimuthal coordinates $r$, $z$, and $\theta$ (with corresponding unit vectors $\hat r$, $\hat z$, and $\hat \theta$). 
If there are no gradients in the $\hat \theta$ direction, and if $\zeta_\theta = 0$, then 
\begin{gather}
\frac{\bB \cdot \nabla \times \bz}{B^3} = \frac{B_\theta}{B^3} \bigg( \frac{\partial \zeta_r}{\partial z} - \frac{\partial \zeta_z}{\partial r} \bigg) . 
\end{gather}
If $B_\theta = 0$, this vanishes, as does the rest of both Eq.~(\ref{eqn:integrandVanishes}) and Eq.~(\ref{eqn:integrandVanishesWithDivergence}), since then all terms in the triple product lie in the $\hat r, \hat z$ plane. 
In other words, the integrand vanishes everywhere for a system with rotational symmetry, so long as $B_\theta$ and $\zeta_\theta$ also vanish. 

\subsection{Generalized $p(B)$ Equilibria}

In the case without $\bz$, perhaps the better-known special equilibria are the $p(B)$ solutions first described by Taylor \cite{Taylor1963, Post1987}. 
The original form of these solutions assumed the low-$\beta$ limit. 
If $p_{||}$ and $p_\perp$ are functions of $B$ alone, then $\nabla( p_{||} + p_\perp)$ is proportional to $\nabla B$, and the integrand of Eq.~(\ref{eqn:CGLPerpLowBeta}) (the second solvability condition) vanishes and Eq.~(\ref{eqn:CGLParallel}) (the first condition) becomes 
\begin{gather}
B \, \frac{\D p_{||}}{\D B} = p_{||}-p_\perp . 
\end{gather}
Taylor showed that one family of low-$\beta$ $p(B)$ solutions is 
\begin{align}
p_{||} &= \begin{cases} C B (B_0 - B)^k & B < B_0 \\ 0 & B \geq B_0 \end{cases} \label{eqn:pBPar} \\
p_\perp &= \begin{cases} k C B^2 (B_0 - B)^{k-1} & B < B_0 \\ 0 & B \geq B_0, \end{cases} \label{eqn:pBPerp}
\end{align}
where $k$, $B_0$, and $C$ parameterize the family of solutions. 
They can be produced by distributions of the form 
\begin{gather}
f(\mu, \epsilon) = \begin{cases} 
(\mu B_0 - \epsilon)^{k-3/2} g(\mu) & \epsilon < \mu B_0 \\
0 & \epsilon \geq \mu B_0 
\end{cases}
\end{gather}
for arbitrary $\mu$ distributions $g(\mu)$. 
Here $\mu \doteq m v_\perp^2 / 2 B$ is the magnetic moment and $\epsilon \doteq m v^2 / 2$ is the energy for particles with mass $m$. 

It was later found \cite{Northrop1964, Taylor1965} that there also exist $p(B)$ solutions for finite $\beta$. 
Ref.~\cite{Northrop1964} predates any calculation of finite-$\beta$ solvability conditions. 
However, if $p_{||}$ and $p_\perp$ are functions of $B$ alone, and if $\bz = 0$, then Eq.~(\ref{eqn:integrandVanishes}) becomes
\begin{gather}
\frac{\D (p_{||}+p_\perp)}{\D B} \, \nabla B \cdot \mathbf{j}_\perp = 0
\end{gather}
and Eq.~(\ref{eqn:integrandVanishesWithDivergence}) becomes 
\begin{gather}
\frac{\D (p_{||}+p_\perp)}{\D B} \, \nabla B \cdot \mathbf{j}_\perp - B^2 \nabla \cdot \bigg[ \frac{(p_{||}-p_\perp) \mathbf{j}_{||}}{B^2} \bigg] = 0 . 
\end{gather}
The equilibria described by Northrop and Whiteman satisfy both of these conditions, though they include boundary conditions that make them somewhat more restrictive. 

The $p(B)$ solutions are interesting because they provide simple examples of allowed equilibria in an arbitrary field, and because they turn out to have excellent stability properties \cite{Taylor1963, Taylor1965, Post1987}. 
A full generalization of these equilibria when $\bz \neq 0$ is difficult in general. 
However, there are some cases which are relatively straightforward. 

Consider the case in which $\bz = - \nabla \Phi$ for some scalar function $\Phi$. 
Then the solvability conditions Eqs.~(\ref{eqn:generalParallelCondition}) and (\ref{eqn:solvability}) -- and, in fact, even the more basic governing equation Eq.~(\ref{eqn:CGLForceBalance}) -- can be transformed back to the $\bz = 0$ case with the substitution 
\begin{align}
p_{||} &\rightarrow p_{||} + \Phi \\
p_\perp &\rightarrow p_\perp + \Phi. 
\end{align}
This makes it possible to translate known equilibria (like the $p(B)$ solutions) with $\bz = 0$ to new solutions with $\bz = - \nabla \Phi$. 
However, this leads to subtleties. 

Consider, for example, the low-$\beta$ solutions given in Eqs.~(\ref{eqn:pBPar}) and (\ref{eqn:pBPerp}). 
To generalize these solutions, consider cases in which $p_{||} + \Phi$ and $p_\perp + \Phi$ are functions of $B$ alone. 
This is computationally straightforward, but note that the analogous boundary condition is now to fix the values of $p_{||} + \Phi$ and $p_\perp + \Phi$ at some given $B$. 
This leads to 
\begin{align}
p_{||} &= \begin{cases} C B (B_0 - B)^k - (\Phi - \Phi_0) & B < B_0 \\ 0 & B \geq B_0 \end{cases} \\
p_\perp &= \begin{cases} k C B^2 (B_0 - B)^{k-1} - (\Phi - \Phi_0) & B < B_0 \\ 0 & B \geq B_0, \end{cases} 
\end{align}
for some $\Phi_0$. 
There are cases in which these may be physically reasonable boundary conditions (for example, if $\Phi-\Phi_0$ and the pressures all vanish on the same surface of constant $B = B_0$, and if $\Phi$ exceeds $\Phi_0$ wherever $B < B_0$), but there is no guarantee that this will be the case. 

The special case in which $\bz$ is the gradient of a scalar potential is \textit{almost} the case of greatest interest. 
However, $\bz$ corresponds to the force density on the plasma (inertial or otherwise), not the force itself. 
Therefore, if the plasma is subject to some potential $\varphi$ (for example, the centrifugal potential due to rotation) then we have instead 
\begin{gather}
\bz = - n \nabla \varphi, \label{eqn:potentialForce}
\end{gather}
where $n \doteq \rho / m$ and $m$ is the ion mass. 
In this case, 
\begin{gather}
\nabla \times \bz = - \nabla n \times \nabla \varphi. 
\end{gather}
In other words, if the gradient of $n$ does not align with the gradient of $\varphi$, then $\bz$ itself will not be a total gradient. 
Then Eq.~(\ref{eqn:solvability}) becomes 
\begin{empheq}[box=\fbox]{align}
&\oint \bigg[ \frac{\nabla (p_{||}+p_\perp) + 2 n \nabla \varphi}{B^3} \cdot (\bB \times \bk) \nonumber \\
&\hspace{90 pt}- \frac{\bB \cdot (\nabla n \times \nabla \varphi)}{B^3} \bigg] \, \D s = 0. 
\end{empheq}
This can be written alternatively as 
\begin{align} 
&\oint \frac{\nabla (p_{||} + p_\perp)}{B^3} \cdot (\bB \times \bk) \, \D s \nonumber \\
&\hspace{10 pt}= \oint \frac{\nabla \varphi}{B^3} \cdot \big[ \bB \times ( \nabla n - 2 n \bk ) \big] \, \D s. 
\end{align}

\section{Isodynamicity} \label{sec:isodynamicity}

The original form of the solvability conditions is closely related to the early theory of omnigeneity and isodynamicity. 
Consider Clebsch coordinates $(\psi, \beta, \chi)$, where $\psi$ labels flux surfaces, 
\begin{gather}
\bB = \nabla \psi \times \nabla \beta, 
\end{gather}
and 
\begin{gather}
B^2 = (\nabla \psi \times \nabla \beta) \cdot \nabla \chi. 
\end{gather}
Let $\bv_d$ denote the cross-field drift velocity. 
Omnigeneity can be understood as the condition that 
\begin{gather}
\langle \bv_d \cdot \nabla \psi \rangle = 0, 
\end{gather}
where $\langle * \rangle$ denotes a bounce-averaging operation. 
A stricter version of this condition is to require instead that 
\begin{gather}
\bv_d \cdot \nabla \psi = 0
\end{gather}
at every point.
This leads to configurations that Catto and Hazeltine \cite{Catto1981} call ``locally omnigenous" configurations, but which are also called ``isodynamic" \cite{Helander2014, RodriguezThesis}. 

Catto and Hazeltine showed that, so long as $\mathbf{j} \cdot \nabla \psi$ vanishes, this local omnigeneity condition is equivalent to the condition that 
\begin{gather}
\nabla \cdot \mathbf{j}_{||} = 0. \label{eqn:vanishingJPar}
\end{gather} 
Their result holds both for the scalar and the anisotropic forms of the pressure tensor considered in this paper (though their force balance included no inertial or other additional force $\bz$). 
Eq.~(\ref{eqn:vanishingJPar}) is equivalent to the condition that the integrand in the second solvability condition -- Eq.~(\ref{eqn:idealMHDPerp}) for the scalar case and Eq.~(\ref{eqn:CGLPerpFiniteBeta}) for the anisotropic case -- must vanish. 
This helps to explain the nice properties of the special equilibria obtained by forcing the solvability integrand to vanish at every point. 

One might have hoped that the generalized form of the solvability integrals (with $\bz \neq 0$) would also lead to a generalization of Catto and Hazeltine's result, yielding a simple condition for isodynamicity in the presence of an arbitrary additional term in the momentum equation. 
Unfortunately, things are not so simple. 

To see why, note that the $\nabla B$ and curvature drifts can be written (in combined form) as 
\begin{gather}
\bv_m = \frac{\hat b}{q B} \times \big( \mu \nabla B + m v_{||}^2 \bk \big), 
\end{gather}
where $\mu = m v_\perp^2 / 2 B$ is the first adiabatic invariant, $q$ is the charge, and $v_{||}$ is the parallel velocity. 
In $(\psi, \beta, \chi)$ coordinates, using the assumption that $\mathbf{j} \cdot \nabla \psi=0$, 
\begin{align}
\bv_m \cdot \nabla \psi &= - \frac{\mu B + m v_{||}^2}{q B} \frac{\partial B}{\partial \beta} \, . \label{eqn:vmCrossField}
\end{align}
Catto and Hazeltine's argument consists essentially of demonstrating in the scalar case that $\nabla \cdot \mathbf{j}_{||} \propto \partial B / \partial \beta$ and in the anisotropic case that $\partial B / \partial \beta = 0$ requires $\nabla \cdot \mathbf{j}_{||} = 0$. 

If for some arbitrary force $\bF$ we write 
\begin{gather}
\bF = F_\psi \nabla \psi + F_\beta \nabla \beta + F_\chi \nabla \chi, 
\end{gather}
and if $\bv_F$ is the $\bF\times\bB$ drift due to force $\bF$, then 
\begin{gather}
\bv_F \cdot \nabla \psi = \frac{F_\beta}{q} \, . \label{eqn:vFCrossField}
\end{gather}
It might be possible to get cancellation between nonzero $\bv_m \cdot \nabla \psi$ and $\bv_F \cdot \nabla \psi$ for a particle with some particular $v_{||}^2$ and $v_\perp^2$, but this is not sufficient. 
In order for the configuration to be isodynamic, the drifts in the $\nabla \psi$ direction must vanish for \textit{all} particles. 
In order for a single condition on the fields to guarantee that $(\bv_m + \bv_F) \cdot \nabla \psi = 0$, $\bv_F$ would have to have a particular form -- for example, either $F_\beta \propto \partial B / \partial \beta$ or $F_\beta \propto \mu B + m v_{||}^2$. 
In general, then, no single solvability condition can be equivalent to the condition that $(\bv_m + \bv_F) \cdot \nabla \psi$ vanish.

\section{Conclusion} \label{sec:conclusion}

This calculation has described the generalization of the MHD solvability conditions for systems with additional forces or steady-state flows at arbitrary $\beta$. 
Solvability conditions are interesting in and of themselves because they provide a general description of the space of possible solutions for a single-fluid model. 
More practically, they can also be used to generate families in equilibria in certain special cases. 
The original, zero-flow solvability conditions also led to unexpected insights related to omnigeneity and isodynamicity \cite{Catto1981}. 
For this reason, it is interesting to see their generalization, even though the solvability conditions do not appear to correspond as closely to the conditions for isodynamicity in the case with flows or other additional forces. 

Along the way, this calculation fills in a technical gap in the literature on anisotropic MHD, unrelated to any issues regarding flows or external forces. 
Taylor's derivation of the anisotropic-single-fluid solvability conditions relied only on the single-fluid momentum equation and Maxwell's equations, but Taylor's calculation was only valid in the low-$\beta$ case \cite{Taylor1963}. 
Hall and McNamara calculated the finite-$\beta$ generalization of Taylor's result, but they did so using a different formalism which brings in assumptions about the dependences of the particle distributions \cite{Hall1975}. 
Their assumptions are well-motivated, but it is still of interest to know whether the finite-$\beta$ generalization could have been calculated more minimally, without requiring anything outside of the single-fluid model. 
Here we have shown that it is indeed possible, albeit at the cost of some relatively lengthy calculations. 

\section*{Acknowledgements}
The authors thank Ian Ochs, Jean-Marcel Rax, Eduardo Rodr\'iguez, and Tal Rubin for helpful conversations. 
This work was supported by ARPA-E Grant No. DE-AR0001554. 
This work was also supported by the DOE Fusion Energy Sciences Postdoctoral Research Program, administered by the Oak Ridge Institute for Science and Education (ORISE) and managed by Oak Ridge Associated Universities (ORAU) under DOE Contract No. DE-SC0014664.

%

\bibliographystyle{apsrev4-2}

\providecommand{\noopsort}[1]{}\providecommand{\singleletter}[1]{#1}%

\end{document}